\documentclass[12pt,preprint,referee]{aastex}

\usepackage[usenames]{color}

\usepackage{graphics}
\usepackage{graphicx}
\usepackage{epstopdf}
\usepackage{url}
\usepackage[stable]{footmisc}
\usepackage{amsmath}
\usepackage{color}

\newcommand{\be}{\begin{equation}}
\newcommand{\ee}{\end{equation}}
\newcommand{\ba}{\begin{eqnarray}}
\newcommand{\ea}{\end{eqnarray}}
\newcommand{\bc}{\begin{center}}
\newcommand{\ec}{\end{center}}
\newcommand{\lsi}{LS~I~+61$^{\circ}$303}
\newcommand{\ls}{LS 5039}

\begin{document}

\title{\textsc{Spectral analysis in orbital/superorbital phase space and hints of superorbital variability in the hard X-rays of \lsi }}
\author{Jian Li\altaffilmark{1}, Diego F. Torres\altaffilmark{1,2}, \& Shu Zhang\altaffilmark{3} }
\altaffiltext{1}{Institute of Space Sciences (IEEC-CSIC),
              Campus UAB,  Torre C5, 2a planta,
              08193 Barcelona, Spain}
\altaffiltext{2}{Instituci\'o Catalana de Recerca i Estudis Avan\c{c}ats (ICREA).}
\altaffiltext{3}{Laboratory for Particle Astrophysics, Institute of High
Energy Physics, Beijing 100049, China. }

\keywords{X-rays: observations, X-ray binaries: individual (\lsi\/)}
\begin{abstract}
 We present INTEGRAL spectral analysis in the orbital/superorbital phase space of \lsi\/. A hard X-ray spectrum with no cutoff is observed at all orbital/superorbital phases.
 The hard X-ray index is found to be uncorrelated with the radio index (non-simultaneously) measured at the same orbital and superorbital phases.
In particular, the absence of an X-ray spectrum softening during the periods of negative radio index does not favor a
simple interpretation of the radio index variations in terms of changes of state in a microquasar.
We uncover hints for the superorbital variability
in the hard X-ray flux,  in phase with the superorbital modulation in soft X-rays.
An orbital phase drift
of radio peak flux and index along the superorbital period is observed in the radio data. We explore
its influence on a previously reported double peak structure of radio orbital lightcurve, posing it as a plausible explanation.

\end{abstract}

\section{Introduction}

 \lsi\/ is a high mass X-ray binary hosting a B0 Ve star with an equatorial outflowing disk.
The companion is orbited by a compact object with an orbital period of
26.496 days (e.g., Casares et al. 2005). \lsi\/ is detected up to GeV and TeV gamma-rays (Albert et al. 2006;
 Abdo et al. 2009; Acciari et al. 2009; Hadasch et al. 2012), and the non-thermal spectral energy distribution is dominated by high
 energy photons.
  Two short ($<$ 0.1 s) and highly luminous ($ > 10^{37}$ erg s$^{-1}$) thermal flares were detected recently
from the direction of \lsi\/ (Barthelmy et al. 2008; Burrows et al. 2012), giving support to the hypothesis that the compact object
in \lsi\ is a (at least internally) highly magnetized
neutron star (Torres et al. 2012; Papitto et al. 2012).

Besides the orbital period of 26.496 days, and other short-timescale variability (see e.g., Paredes et al. 2007), \lsi\ is known
to have a long-term (1667 days) superorbital modulation. The latter was first
 discovered in radio (Paredes 1987; Gregory 1989; Gregory 1999; Gregory 2002), and then observed in H${\alpha}$ emission line
 (Zamanov et al. 1999), X-rays (Li et al. 2012;
 Chernyakova et al. 2012), GeV (Ackermann et al. 2013), and hinted at TeV (Li et al. 2012; Torres et al. 2012).
The possible origin of this 1667 days superorbital modulation could be the precession of the Be disk
(Lipunov \& Nazin 1994), the beat frequency between the orbital and
 precessional rates in a microquasar scenario (Massi \& Jaron 2013), or the cyclic variability in the Be star envelope.
 The last interpretation seems the most natural, since the
H${\alpha}$
 emission line varies on the same period and it is well known for the Be stars that the size of
 the circumstellar disk grows as the equivalent width of H${\alpha}$  increases
 (e.g., Hanushik, Kozok \& Kaizer, 1988). In the maximum of the equivalent width of H${\alpha}$, the
X-ray as well as  $\gamma$-ray emission are enhanced
 (Li et al. 2012; Ackermann et al. 2013). The orbital phase of the X-ray peaks from \lsi\ varies from 0.35 to 0.75
along the superorbital period, consistently leading the radio peaks by 0.2 orbital phase (Chernyakova et al. 2012). The different
origin regions of X-ray and radio emission could lead to the observed phase lag.
The observed multiwavelength variabilities in
 superorbital timescale were discussed in the scenario of a neutron star switching between ejector
 and propeller states by Torres et al. (2012) and Papitto et al. (2012).

 Black hole composed microquasars manifest themselves in five classical spectral X-ray states  (van der Klis 1994;
 Tanaka \& Lewin 1995; Tanaka \& Shibazaki 1996; Esin et al. 1997): the high/soft (HS) state,
 the low/hard (LH) state, the quiescent state, the very high (VH) state, and the intermediate state.
 These
 spectral states were regrouped by McClintock \& Remillard (2006) as thermal--dominant (TD),
 hard X-ray, quiescent, steep powerlaw (SPL), and intermediate states.
 From an observational perspective, an optically thick steady jet with flat
 (radio spectral index $\alpha \sim 0$) or inverted ($\alpha \geqslant 0$) radio component is usually
 seen when in the hard X-ray state (a power law spectrum with index 1.5 $< \Gamma <$ 2.1 in X-rays,
 McClintock \& Remillard 2006), and an optically thin transient jet ($\alpha < 0$) is associated with the overall hard to soft
 state transition (Fender et al. 2004; Fender et al. 2009).

In the hypothesis of \lsi\ being a microquasar, the explanation for the short thermal flares that have been detected from the region of \lsi\ must be
the spatial superposition of a magnetar with a gamma-ray binary, however unlikely. In this hypothesis,
Zimmermann \& Massi (2012) expected a steady jet ($\alpha \geqslant 0$)
in the hard X-ray state of the source, thus characterized by a power law with index 1.5 $< \Gamma <$ 1.8 and a cutoff
at high energies. A
 transient jet (with $\alpha < 0$) would on the contrary be expected in the SPL state of \lsi,
 showing an X-ray slope of $ \Gamma \geq $ 2.4.
These seem very simplified assumptions, given that such an SPL state is
 poorly characterized. McClintock \& Remillard (2006) note that many BH
binaries could be radio quiet during this phase, and moreover, such state
 seems often described by high X-ray luminosity ($> 0.2$ times the
Eddington level) and the detection of a thermal component or QPOs; all of which are
clearly not the case of \lsi\ (Li et al. 2011).

Although GBI radio data showed spectral index transitions
(Massi \& Kaufman Bernad\'{o}, 2009), the photon index of INTEGRAL observations was found to be
compatible with
a low hard state (Chernyakova et al. 2006; Zhang et al. 2010) at all times. Only in one occasion the photon
index was $\Gamma=3.6^{+1.6}_{-1.1}$ (Chernyakova et al. 2006).
However, the large error bars (due to little exposure) introduced a large uncertainty.
Zimmermann \& Massi
(2012) demonstrated that averaging the INTEGRAL data over too large orbital and superorbital
phase ranges can result in a dominant low hard state and smear out the SPL state. This effect, they posed, is what explains
the hard spectra all along the orbit in the \lsi\ system, and which led to the
inconsistence between data and their predictions.
Here, using INTEGRAL observations that double the size of our previous analysis (Zhang et al. 2010) and two years more data than in Chernyakova
et al. (2012) we can separate in orbital and superorbital
phases and test these ideas.

\section{Observations and data analysis}

INTEGRAL (Winkler et al. 2003) is a $\gamma$-ray mission covering the energy the range 15
keV--10 MeV. Observations are carried out in individual Science Windows
(ScW), which have a typical time duration of about 2000 s. We use all public IBIS/ISGRI data for which \lsi\ has
an offset angle less than $14{\degr}$. Our data set comprises about 2006 ScWs.
The data cover revolutions 6--1317, from 2002-11-01 to 2013-07-27 (MJD 52579--56500),
adding up to a total effective exposure time of 928 ks in IBIS/ISGRI. The data reduction is
performed using the standard ISDC offline scientific analysis (OSA 10.0).
IBIS/ISGRI images for each ScW are generated in the energy band of 18--60
keV. The count rate at the position of the source are
extracted from all individual images to produce the long-term lightcurve on the ScW timescale.
The spectra of \lsi\ are produced for each of the ScW following the standard steps as stated in the IBIS Analysis User Manual,
running the pipeline from the raw data to SPE level\footnote{Please see \url{http://www.isdc.unige.ch/integral
/analysis} for more information}. All of the spectral analysis is
performed using XSPEC  12.8.1; uncertainties are given at the 1$\sigma$
confidence level for one single parameter of interest.

We have also considered radio data from NASA/NRAO Green Bank Interferometer (GBI) database. The radio data set,  the same as the one used by
Massi \& Kaufman Bernad\'{o} (2009) and Zimmermann \& Massi (2012), covers \lsi\ at $\nu_{1}=$2.25 GHz and $\nu_{2}=$
8.3 GHz for 6.7 years in three periods of MJD 49379.975--50174.710, MJD 50410.044--51664.879 and MJD 51798.333--51823.441.
The radio spectral index $\alpha$ and its error  $\Delta \alpha$  are calculated as
$\alpha={\log(S_1/S_2)\over \log(\nu_1/\nu_2)}$ and $\Delta \alpha={0.434\over \log (\nu_1/\nu_2)}\sqrt{( {\Delta S_1\over S_1})^2+
({\Delta S_2\over S_2})^2}$, where $S_{1,2}$ are the corresponding radio fluxes at the two frequencies of interest.

\section{Results}
\subsection{INTEGRAL spectral analysis in the orbital/superorbital phase space}

The orbital phase of an X-ray binary system represents a unique location of the compact object in the orbit.
The compact object will encounter a similar physical environment (e.g. magnetic field, mass density,
accretion rate, etc.) in the same orbital phase if the properties of companion star are
stable, which will presumably lead to a repetitive pattern of orbital emissions.

Long-term stability of the X-ray orbital lightcurve is observed in \ls\
 (Kishishita et al. 2009).
  Unlike \ls, hosting an O star
  with a relative stable stellar wind, \lsi\/'s companion is a Be star with
  circumstellar disk variability, which presumably gives rise to the superorbital period.
  Because of the different configuration of the
  circumstellar disk along the superorbital period, the physical environment in each orbit is different
  even for the same orbital phase.
  Consequently, we observe variable orbital emission from \lsi\ along
  the superorbital period in all frequencies (Paredes 1987; Gregory 1989; Torres et al. 2010; Li et al. 2012;
  Chernyakova et al. 2012; Ackermann et al. 2013). Averaging data over too large orbital and superorbital
  phase ranges mixes different physical conditions and smears out information (Zimmermann \& Massi 2012).
  In order to expect a similar physical configuration in the \lsi\ system, one should then require
  not only a similar orbital phase, but also a similar superorbital phase.

  With all INTEGRAL data combined, \lsi\ is detected with a significance of 10.59 $\sigma$ in the 18--60 keV band (Fig. 1, left panel).
  INTEGRAL hard X-rays and GBI radio data distribution in the orbital/superorbital phase space are shown in Fig. 2, left panels.
This is similar to  Fig. 4 in Gregory 2002 and
Fig. 2 in Chernyakova et al. 2012 for their respective datasets.
  The number of observations at each position in the phase space is given by the color scale. Although the radio (MJD 49379.975--51823.441) and
  hard X-ray data (MJD 52579--56500) do not overlap in time, some are co-located
  at the same orbital and superorbital phases.
  Fig. 2, upper right
  panel, shows the overlapped INTEGRAL and GBI observations in phase space. The radio index for the phase
  space
  when hard X-rays and radio observations are both available (Fig. 2, left panels) is shown in the
  bottom right panel of the same figure (light blue
stands for $\alpha < 0$; dark blue stands for $\alpha > 0$).

\lsi\ should be in SPL state/low hard state with photon index
  $ \Gamma >$ 2.4 and 1.5 $ <\Gamma <$ 1.8 respectively according to Zimmermann \& Massi (2012).
  To test this, we have extracted the INTEGRAL spectra in the expected SPL state ($\alpha < 0$, light blue region) and low hard state
  ($\alpha > 0$, dark blue region), respectively.
  X-ray spectra are well fitted with a simple power-law without high energy cutoffs. The results are shown
  in Table 1. The X-ray photon index is hard ($ \Gamma $=1.49$^{+0.21}_{-0.19}$)
  when the radio index $\alpha < 0$, what happens for most of the observations.
 This is not consistent with the predictions mentioned. In the case of radio
 index $\alpha > 0$, the X-ray photon index results in $ \Gamma $=2.59$^{+1.01}_{-0.83}$. The large error bars are expected
 from the much less effective exposure (see Table 1). Because of the large uncertainty, no significant difference could be drawn between the two X-ray spectra.

\begin{figure*}[t]
\centering
\includegraphics[angle=0, scale=0.45] {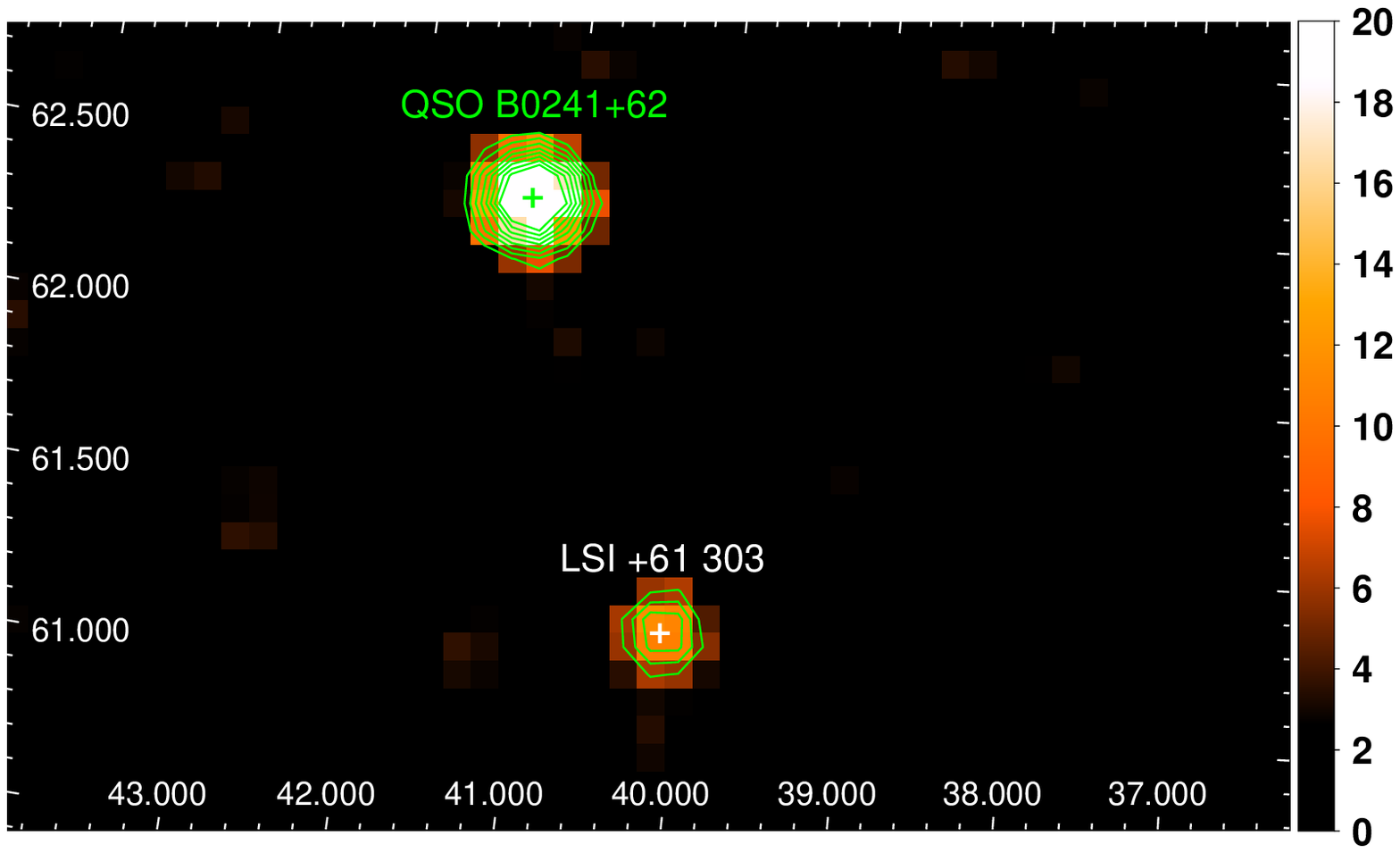}
\includegraphics[angle=0, scale=0.385] {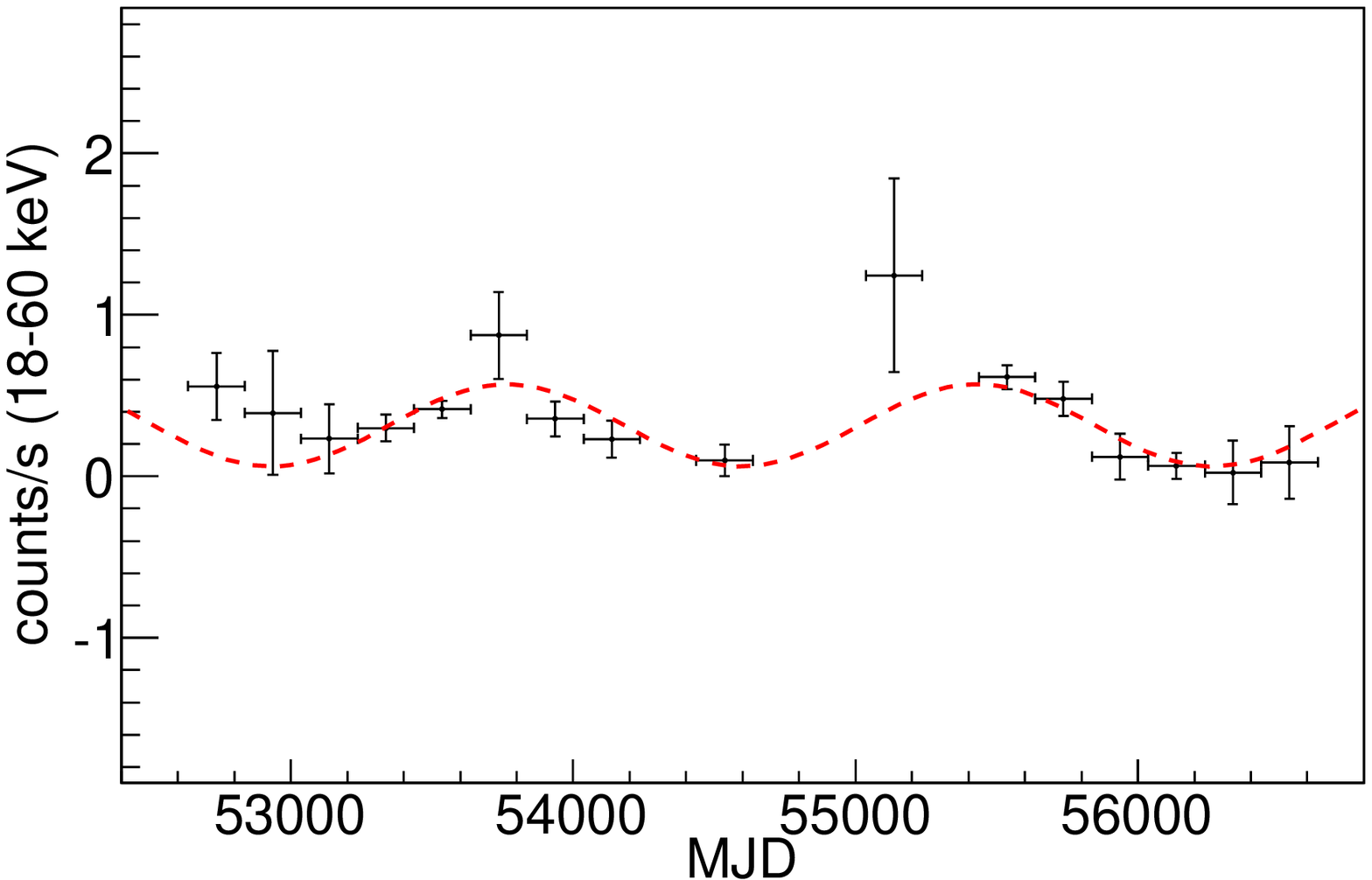}
\caption{Left panel: mosaic image of the \lsi\ sky region, derived by combining all
INTEGRAL/ISGRI data. The brighter source is QSO 0241+622 (at the upper part of the image)
and the relatively faint one is \lsi\/. Corresponding significances and color can be found in
the right color bar. The contours start at a detection significance level of 6 $\sigma$,
with each step being 2 $\sigma$. The X- and Y-axes are R.A. and decl. in units of degrees.
Right panel: long-term lightcurve of \lsi\ in 18-60 keV binned in 200 days. The dotted red line
indicates a sinusoidal fitting with the superorbital period fixed. }
\label{mos-1}
\end{figure*}

 \begin{figure*}[t]
\centering
\includegraphics[angle=0, scale=0.4082] {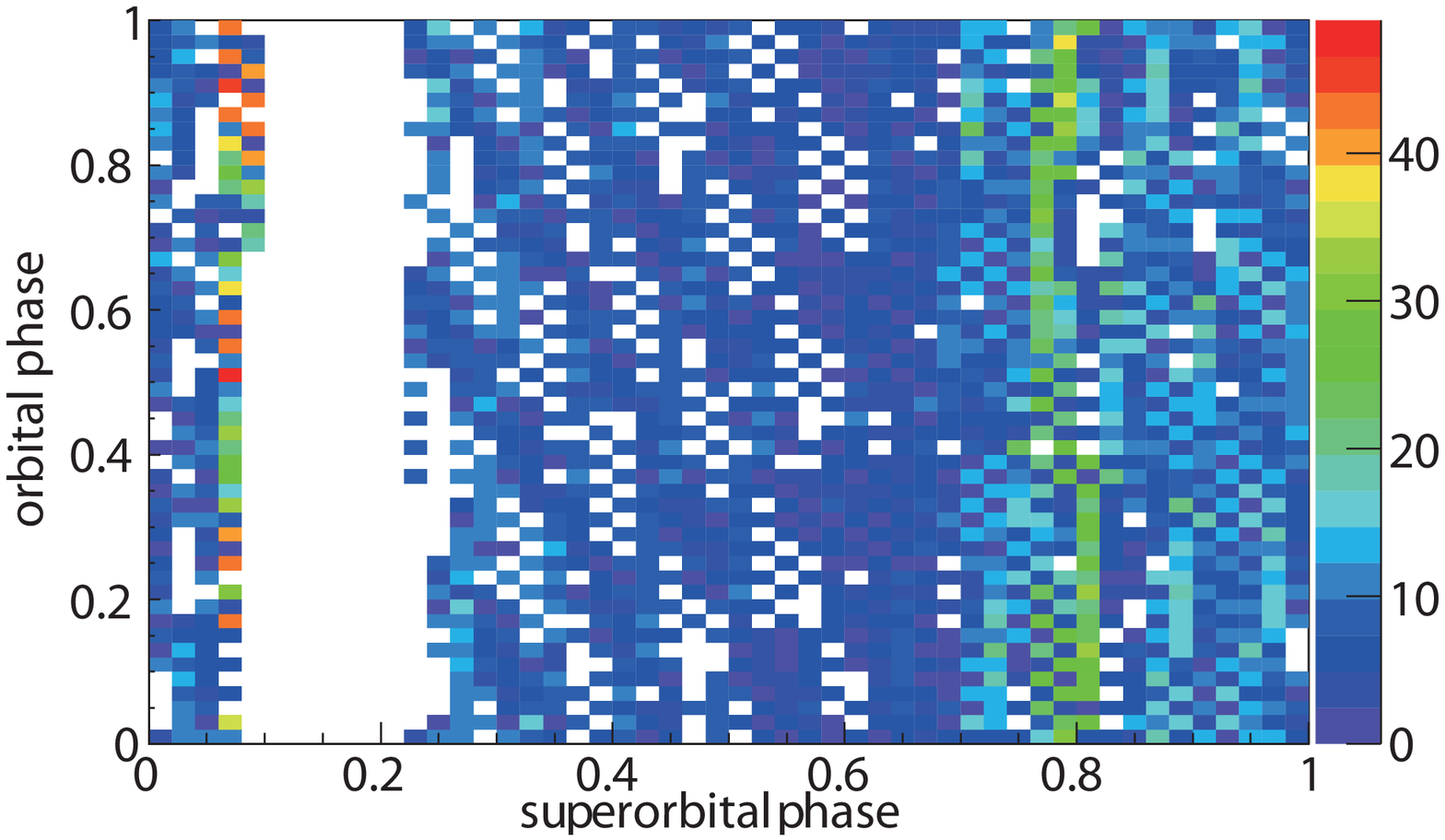}
\includegraphics[angle=0, scale=0.4082] {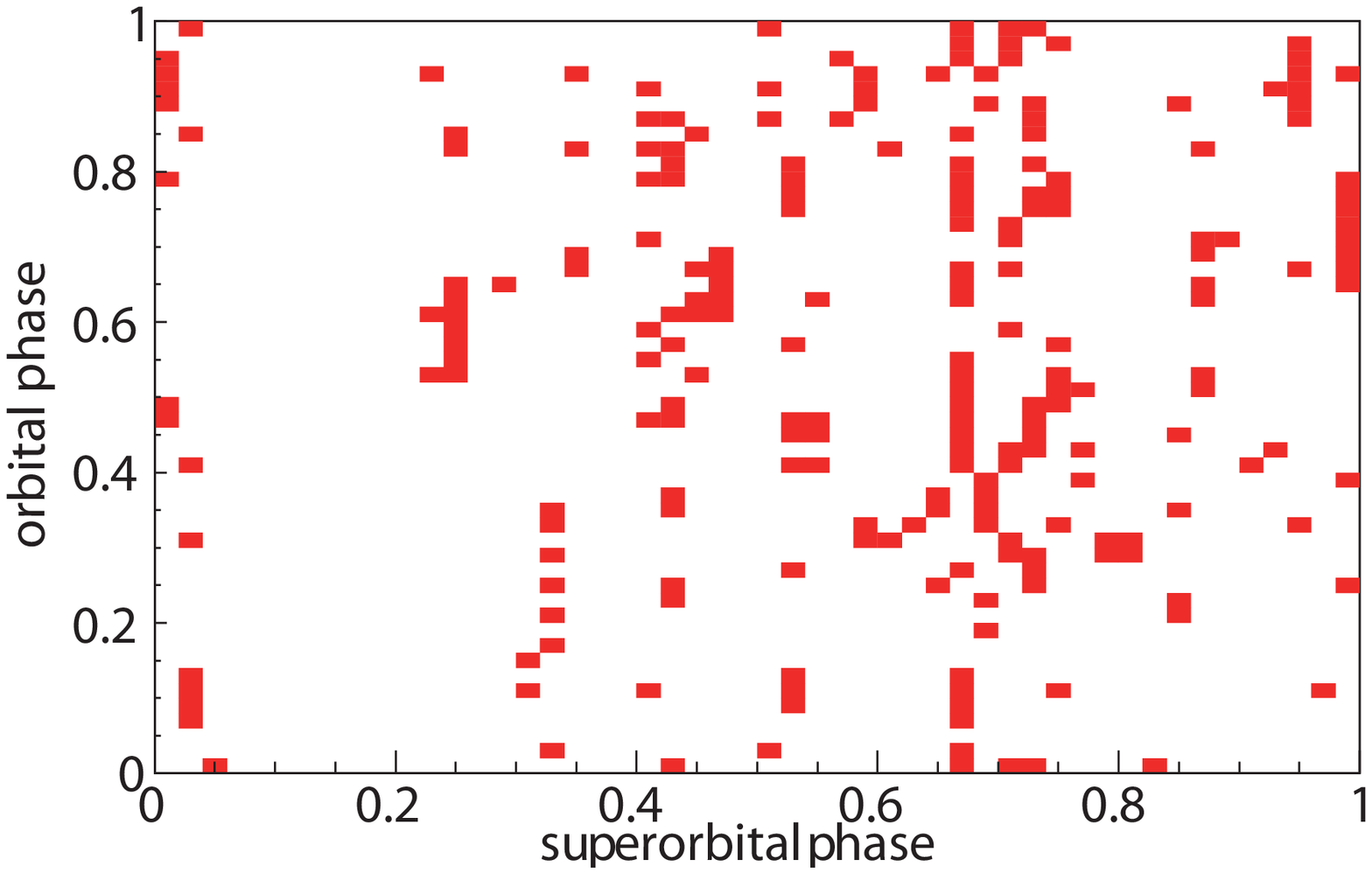}
\includegraphics[angle=0, scale=0.4082] {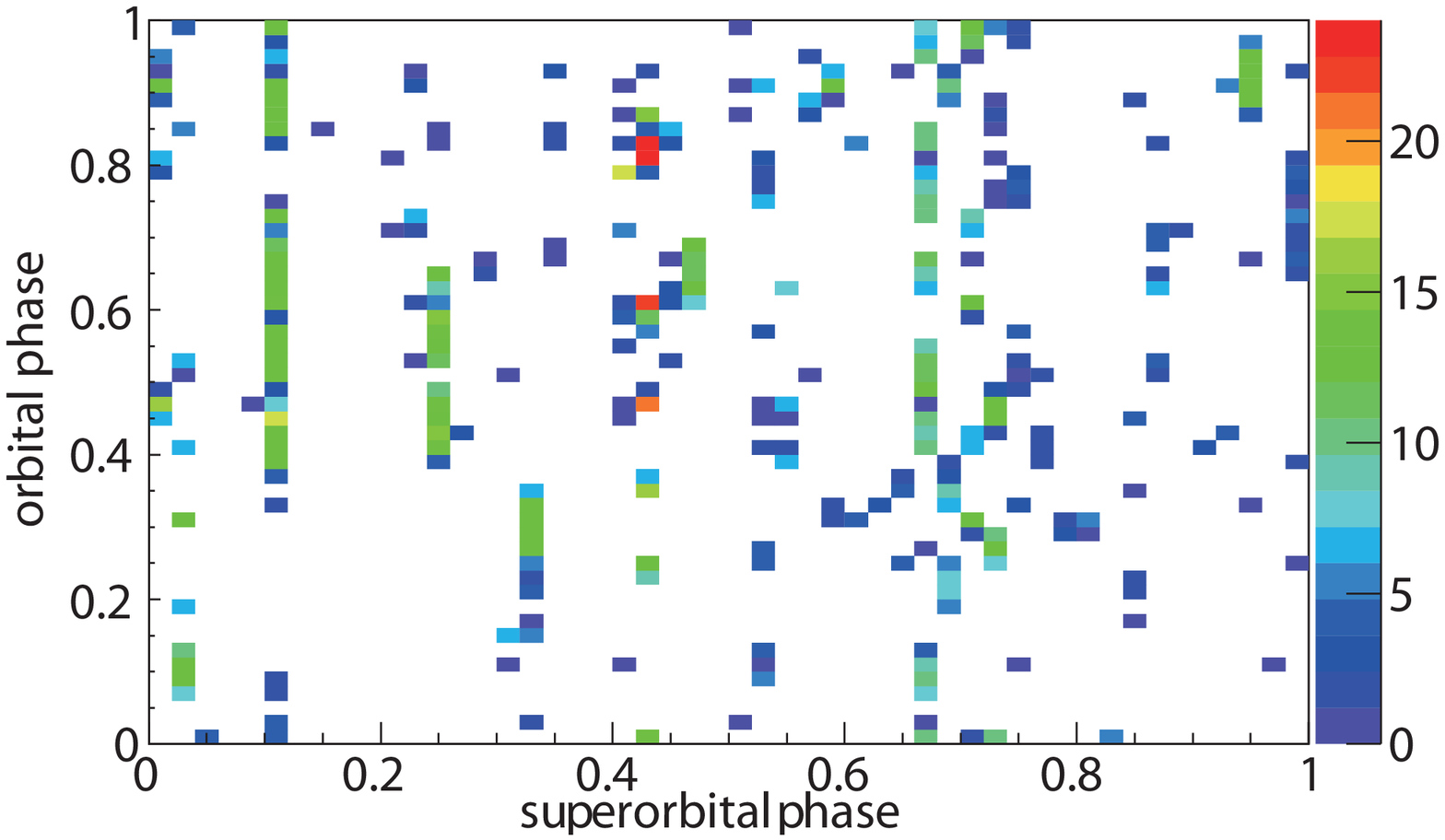}
\includegraphics[angle=0, scale=0.4082] {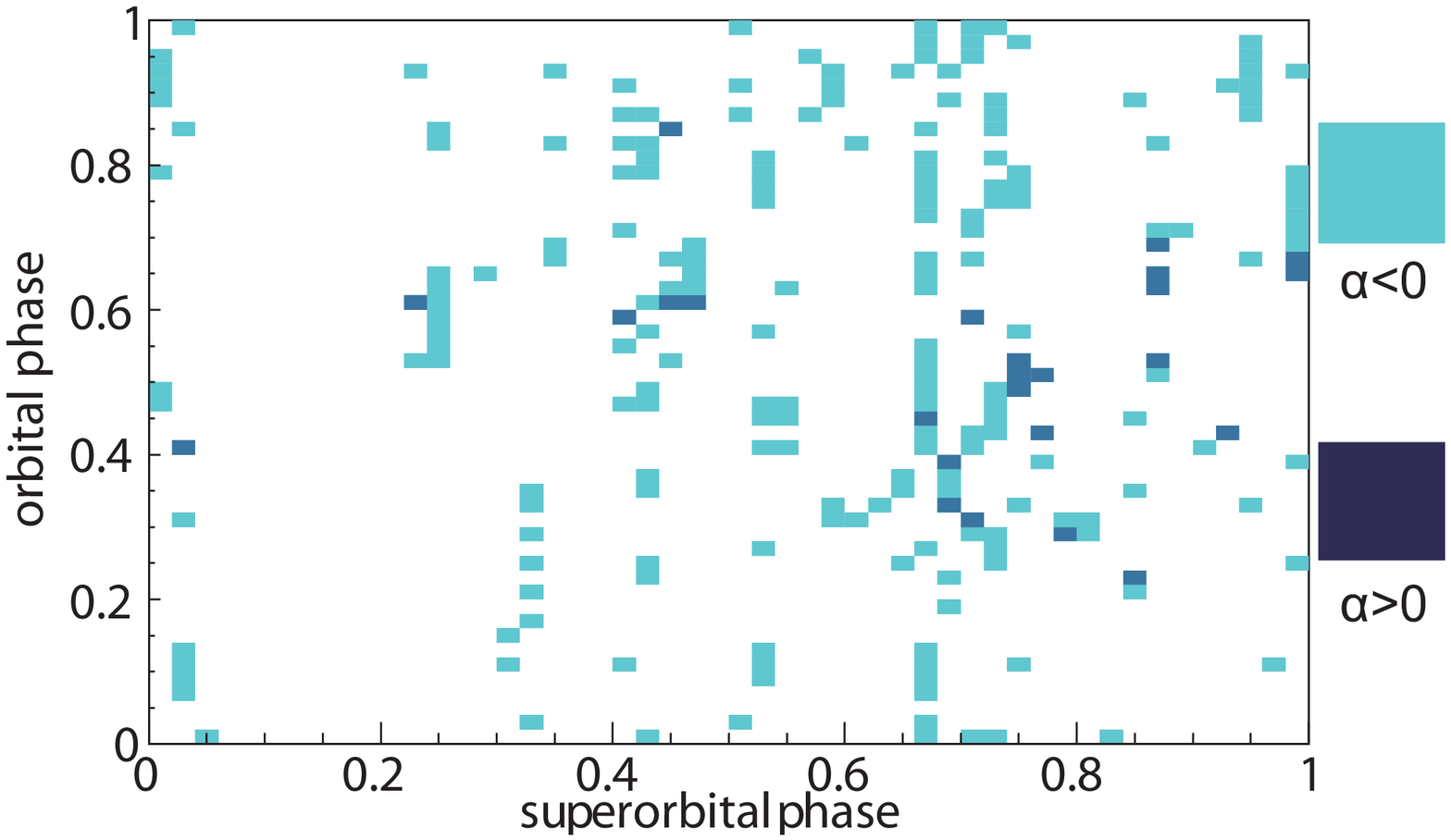}
\caption{GBI (upper left) and INTEGRAL (bottom left) data distribution in orbital/superorbital phase space.
The number of observations are given by the color scale. The upper right panel shows the part of the
phase space with overlapped GBI and INTEGRAL observations. The radio
index of overlapped INTEGRAL and GBI observations are shown in the bottom right panel (light blue
stands for $\alpha < 0$; dark blue stands for $\alpha > 0$). }
\label{mos-1}
\end{figure*}

\subsection{Hints of superorbital variability of \lsi\ in hard X-ray}

The INTEGRAL observations cover \lsi\ for more than 10 years. The long-term lightcurve in the 18--60 keV band
  binned in 200 days is shown in Fig. 1, right panel.  A constant fit to the lightcurve yields an average flux of 0.34$\pm$0.03 counts/s and
a reduced $\chi^{2}$ of 50.92/16, which indicates variability at the 4.3 $\sigma$ level.
The lightcurve can be fitted
by a sinusoidal function with fixed period at 1667 days, yielding a reduced $\chi^{2}$ of 13.42/14 (Fig. 1, right panel).
An F-test shows that the possibility of wrongly refusing the sinusoid is 8.83$\times$10$^{-5}$.

We fold the INTEGRAL ScW lightcurve at 1667 days superorbital period with T$_{0}$ at MJD 43366.275.
A clear modulation profile is seen (Fig. 3, left top panel). A constant fit to the lightcurve yields a reduced $\chi^{2}$ of 36.8/7, indicating variability at 4.6 $\sigma$ level.
The superorbital lightcurve of \lsi\ in
hard X-rays peaks around superorbital phase $\sim$ 0.2, which is in phase with superorbital modulation
in soft X-ray (the red points in Fig. 3, left top panel, showing the modulated fraction in the 3--30 keV band, from Li et al. 2012).
Exposure time and significance of corresponding superorbital phase are shown in Fig. 3 (left), middle and
bottom panels. The large error bar of the superorbital lightcurve peak happens as a result of the low
exposure time ($\sim$7.4 ks) it has compared to other phases (hundreds ks).
We extract spectra for different
superorbital and orbital phases. All spectra are well fitted by a simple powerlaw. The fitting parameters are shown
in Table 1.


  \begin{figure*}[t!]
\centering
\includegraphics[angle=0, scale=0.49] {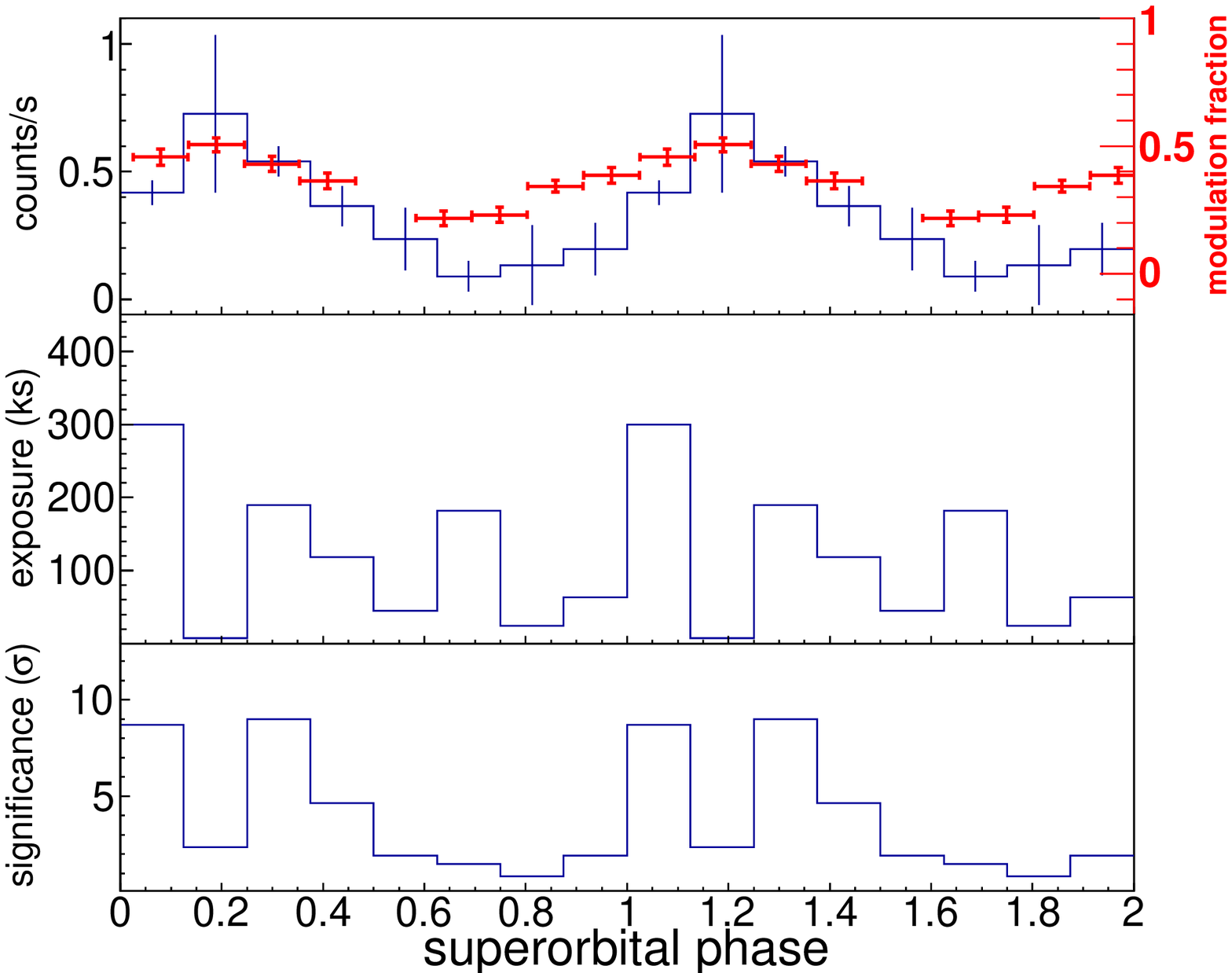} \hspace{0.04cm}
\includegraphics[angle=0, scale=0.31] {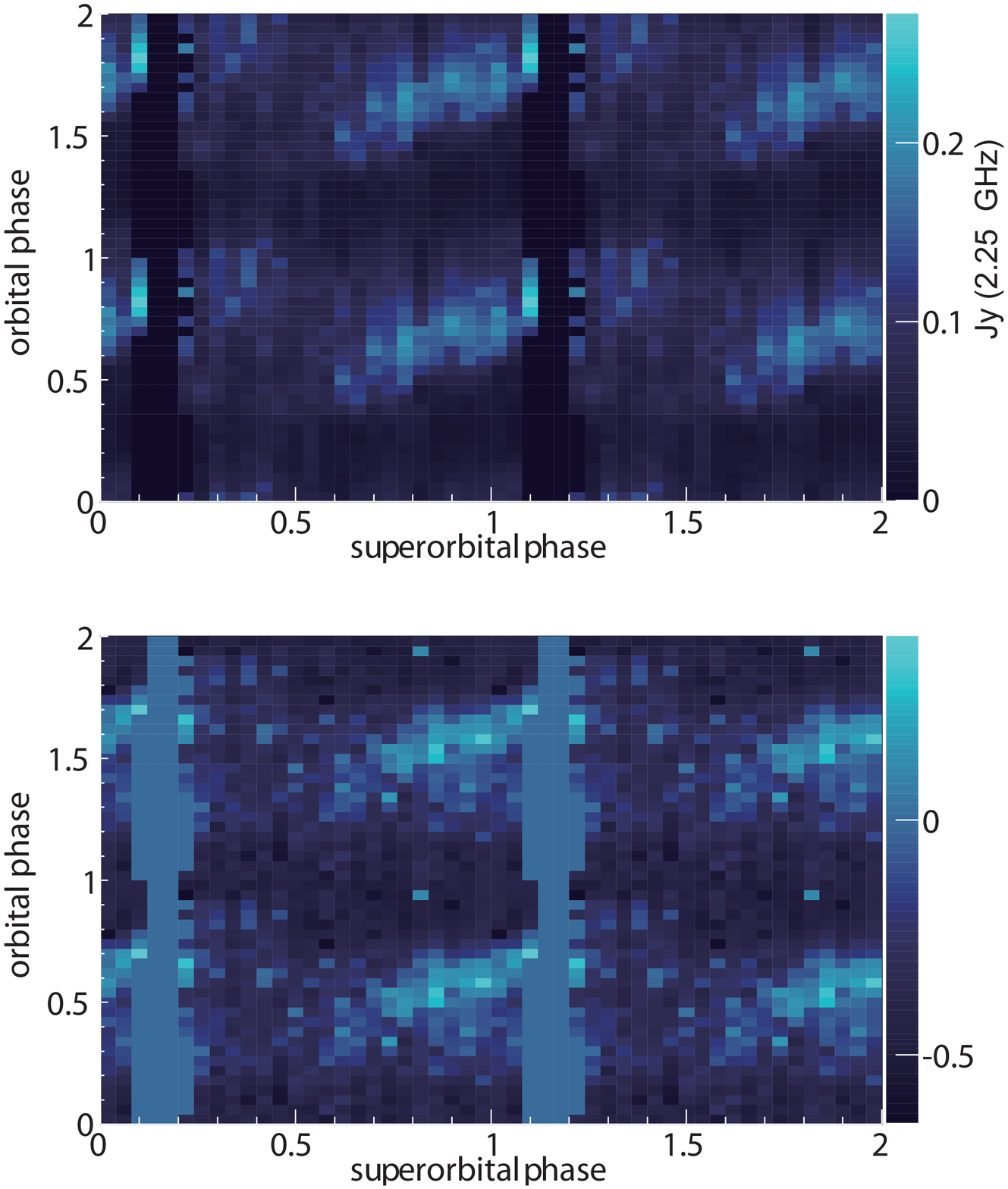}
\caption{Left: Superorbital lightcurve of \lsi\ in the 18--60 keV, as observed by INTEGRAL. The red points show the modulated fraction in the 3--30 keV band, from Li et al. 2012.
Middle and bottom: exposure and significance for the corresponding superorbital phase, respectively. Right: radio flux intensity at 2.25 GHz (top) and spectral index (bottom) as a function of the orbital and superorbital phases.}
\label{mos-1}
\end{figure*}

\begin{table*}[t!]
\centering
\scriptsize
\caption{X-ray spectra parameters associated with observations providing a
radio index $\alpha < 0$ and $\alpha > 0$ and superorbitally / orbitally separated spectra from INTEGRAL/ISGRI observations. }
\vspace{0.1cm}       
\label{table1}      
\centering                          
\begin{tabular}{ccccc}        
\hline\hline                 
\\
Radio index & X-ray photon index &  Flux (18--60 keV)  & Reduced $\chi^{2} (D.O.F)$& Effective exposure\\
             & $ \Gamma $         &  10$^{-11}$ erg cm$^{-2}$ s$^{-1}$ &      &  ks\\
\\
\hline
\hline
\\
$\alpha < 0$ & 1.49$^{+0.21}_{-0.19}$  & 1.67$\pm$0.19  &  0.86 (4) & 508.0\\
\\
\hline
\\
$\alpha > 0$ & 2.59$^{+1.01}_{-0.83}$ & 2.10$^{+0.58}_{-0.59}$  & 0.57 (5)& 42.8\\

\\
\hline\hline                 
\\
Superorbital phase & X-ray photon index &  Flux (18-60 keV)                   & Reduced $\chi^{2} $(D.O.F)& Effective exposure\\
                   & $ \Gamma $         &  10$^{-11}$ erg cm$^{-2}$ s$^{-1}$ &                          &  ks \\
\\
\hline
\hline
\\
0.1-0.2 & 1.06$_{-0.23}^{+0.22}$  &2.14$\pm$0.23   & 0.254 (5)& 238.9 \\
\\
\hline
\\
0.2-0.3 & 1.79$_{-0.21}^{+0.23}$  &3.65$\pm$0.36    &0.330 (7)& 133.1 \\
\\
\hline
\\
0.3-0.5 & 1.66$_{-0.28}^{+0.31}$  &1.87$\pm$0.29   & 1.00 (4)& 182.3\\
\\
\hline
\\
0.5-0.8 & 1.27$_{-0.38}^{+0.40}$  & 1.31$\pm$0.29   & 0.975 (3)& 241.0\\
\\
\hline
\\
0.8-1.1 & 1.27$_{-0.39}^{+0.44}$  & 1.52$_{-0.40}^{+0.42}$  & 0.688 (4)& 134.4\\
\\
\hline
\hline
\\                            
Orbital phase & X-ray photon index &  Flux (18-60 keV)                   & Reduced $\chi^{2} $(D.O.F)& Effective exposure\\
                   & $ \Gamma $         &  10$^{-11}$ erg cm$^{-2}$ s$^{-1}$ &                     & ks\\
\\
\hline
\hline
\\
0.0-0.4 & 1.90$_{-0.40}^{+0.47}$  &1.63$\pm$0.29   & 0.256 (4)& 219.3 \\
\\
\hline
\\
0.4-0.5 & 1.60$_{-0.21}^{+0.22}$  &2.84$\pm$0.30    &0.763 (5)& 162.3 \\
\\
\hline
\\
0.5-0.6 & 1.54$_{-0.25}^{+0.28}$  &2.84$\pm$0.39   & 0.358 (8)& 118.2\\
\\
\hline
\\
0.6-0.7 & 1.44$_{-0.30}^{+0.32}$  & 2.24$\pm$0.35   & 0.424 (4)& 154.2\\
\\
\hline
\\
0.7-1.0 & 0.70$_{-0.34}^{+0.33}$  & 0.98$_{-0.27}^{+0.28}$  & 0.26 (3)& 273.7 \\
\\
\hline
\\                            
\end{tabular}
\end{table*}

\subsection{Orbital phase drift of the radio peak flux and index }

In X-ray and radio bands, the superorbital modulation manifest itself not only in the peak flux or the modulated
  fraction variation (Gregory 2002; Li et al. 2012), but also in a systematic drift of the orbital phase of
  peak flux and spectral index (Ray et al. 1997; Gregory 2002; Chernyakova et al. 2012).
  We have detected the drift of the peak
  flux and spectral index in radio data, consistent with the results in Gregory (2002) and
  Chernyakova et al. (2012).
  From superorbital  phase $\sim$ 0.6 to $\sim$ 1.5,
  the orbital phase of the radio peak flux (both 2.25 GHz and 8.3 GHz data show
  a similar evolution, see Fig. 3)
  is linearly drifting from orbital phase $\sim$ 0.5 to $\sim$ 0.95.
  The linear drift of the spectral index
is less evident, but clearly shifted earlier --by $\sim$ 0.1 in orbital phase-- compared to the flux (Fig 3, right bottom panel).
   Because of the scarce orbital coverage of INTEGRAL along the superorbital period (Fig. 2, bottom
  left panel), we are unable to explore the presence of a peak flux drift in hard X-rays yet.

  Massi \& Kaufman Bernad\'{o} (2009) claimed that the periodic radio orbital outbursts consisted of two peaks.
We pose here that the two-peak structure is related to the drift of radio peak flux along the
superorbital period. Fig. 3 of Massi \& Kaufman Bernad\'{o} (2009) shows the radio orbital
lightcurve during superorbital phase 0.0--0.1, in which the authors claimed two peaks at orbital phase
0.69 and 0.82.
But in our Fig. 3, right top panel, it is clear that the orbital phase of radio peaks drifts from $\sim$0.7 to $\sim$0.8
during superorbital phase 0.0--0.1.
Thus, to explore drifts in detail, we display the lightcurves of 2.25 GHz (top panel), 8.3 GHz (middle panel) and the radio index (bottom panel)
during superorbital phase 0.0--0.1 in Fig. 4, where each orbit is noted
with a different color. The radio data cover more than one superorbital period, from superorbital phases $\sim$3.6 to $\sim$5.1;
the data in  the superorbitally folded bin 0.0--0.1 was gathered in two passages corresponding
to superorbital phase 4.0--4.1 and 5.0--5.1. These results are shown in Fig. 4.
In the right panel of Fig. 4 we show the corresponding folded orbital lightcurve
in superorbital phase 0.0--0.1, similarly to Fig. 3 of Massi \& Kaufman Bernad\'{o} (2009), but keeping
the different color for the different orbits.
The phases of the reported two peaks
(Massi \& Kaufman Bernad\'{o} 2009)
are indicated with dashed black lines.
In Fig. 4, right top and middle panel, it is apparent to the
eye that the phase of peak flux in 2.25 GHz and 8.3 GHz is drifting in each orbit. The orbital
lightcurve depicted in green is peaking around orbital phase $\sim$ 0.69, which is the
first peak reported in Massi \& Kaufman Bernad\'{o} (2009), while several orbits later, the black-colored lightcurve
is peaking at orbital phase $\sim$ 0.82, which is near the second peak reported.
Thus, folded in superorbital phase (0.0--0.1),
the two peaks in Fig. 3 of Massi \& Kaufman Bernad\'{o} (2009)  may actually be a
superposition of single peaks from different orbits.
Similarly, this orbital peak drifting may also have effects in Fig. 5 of Massi \& Kaufman Bernad\'{o} (2009),
where the data is also folded over a bin of 0.1 superorbital
phase. Thus, only the lightcurve of each individual orbit, or the orbital lightcurve folded in a narrower
superorbital phase that avoids the
orbital peak-drifting effect is suitable to explore the reality of the double peak structure, which is here put in question.

\section{Discussion}

 We have tested one of the methods proposed to identify the microquasar nature of the \lsi\ system: Namely, that it is in SPL state when the radio
index $\alpha < 0$ and in low hard state when $\alpha > 0$, and that using large orbital and superorbital phase ranges mixes these
states smearing out the state transitions signatures (Zimmermann \& Massi 2012). The latter has been claimed as the reason of the non-detection of significant
spectral variations in the INTEGRAL data analyzed so far.
Here, with a significantly increased dataset,  we have introduced the orbital / superorbital phase
space and extracted the X-ray spectrum for the corresponding specific cuts,
avoiding averaging data over a too large superorbital range.
\lsi\ shows a hard spectrum (single power law with no cutoff,
$ \Gamma $=1.49$^{+0.21}_{-0.19}$, see Table 1) when $\alpha < 0$, which is
inconsistent with the SPL state interpretation of
Zimmermann \& Massi (2012) stated above.

 As commented in the introduction, such a clear cut distinction between states based on the radio index seems unlikely.
The SPL state itself is not an ``intermediate state"  that lies between soft and hard states,
though transitions pass through SPL frequently (McClintock \& Remillard 2006; Zhang et al. 2013).
It may also be unrealistic to think that
optically thin radio emission is confined to the SPL state only.
If a hard to soft state transition does not pass through an SPL state (e.g., the 2011 outburst of H 1743-322
during which the photon index is always below 2.4, Zhou et al. 2012) or optically thin radio
flares appear in other intermediate states but not in SPL state, the X-ray photon index will
not go above 2.4 during optically thin radio emission. Thus, the X-ray spectrum during
$\alpha < 0$ might be a mixture of a
SPL state or transitions to SPL states
(e.g. see the cases of XTE J1650-500, Corbel et al. 2004; GX 339-4, Gallo et al. 2004; XTE J1859+226, Brocksopp et al. 2002,
Corbel et al. 2004), and intermediate states when optically thin radio emissions locate out of SPL
states. Nevertheless, the fact that most of the observation time in INTEGRAL indeed has a concurrent $\alpha < 0$
(at the same orbital and superorbital phase) indicates that the most likely alternative is simply that there is no state transition at all,
something also supported by the low X-ray flux at all times.

\begin{figure*}[t!]
\centering
\includegraphics[angle=0, scale=0.8] {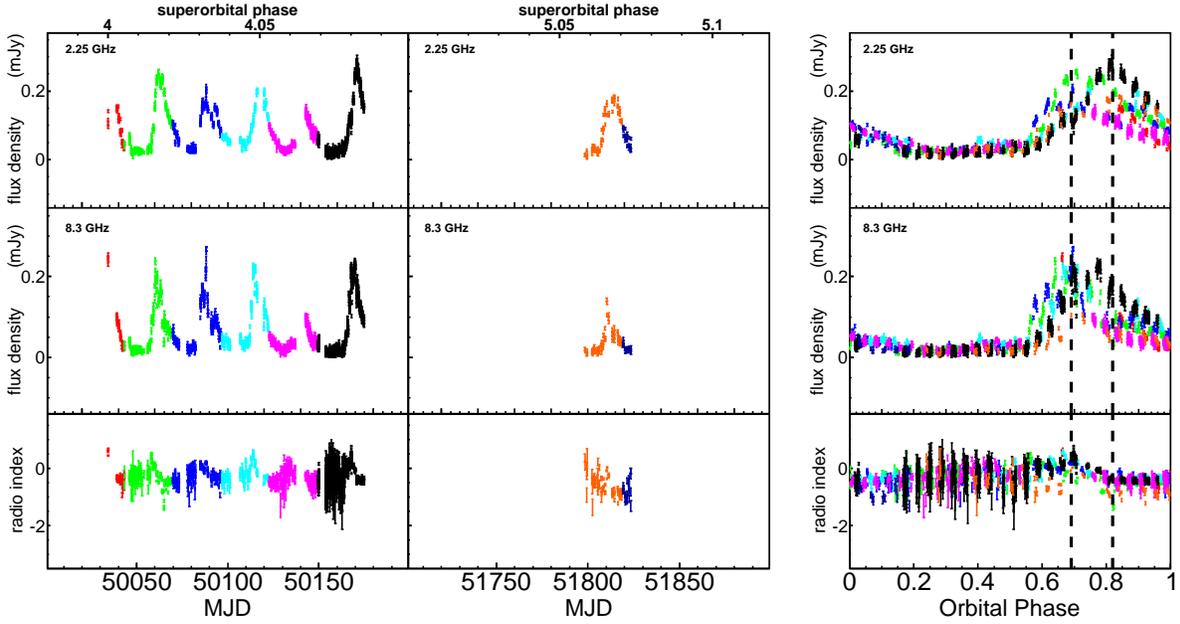}
\caption{Left {and Middle} Panel: lightcurves of 2.25 GHz (top panel), 8.3 GHz (middle panel) and radio index
(bottom panel) during superorbital phase 0.0--0.1. {Superorbital phase is shown as the upper axis on top of the panels}.
Right Panel: folded orbital lightcurve of
2.25 GHz (top panel), 8.3 GHz (middle panel) and radio index (bottom panel) during
superorbital phase 0.0--0.1. The two peaks reported in Massi \& Kaufman Bernad\'{o} 2009 are shown with
dashed black lines. Different orbit (orbital
phase 0.0--1.0) are shown with different color.}
\label{mos-1}
\end{figure*}

We have also shown hints for the superorbital variability of \lsi\ in hard X-rays,
and that the
appearance of a double peak structure in the radio lightcurve of superobitally folded data can be accommodated by a drift of a single peak in an orbital basis.

\acknowledgements

We acknowledge the grants AYA2012-39303,
as SGR2009-811, iLINK2011-0303, NSFC 11073021, 11133002, 11103020, XTP project XDA04060604 and the Strategic Priority Research Program ``The Emergence of Cosmological Structures" of the Chinese Academy of Sciences, grant No. XDB09000000.
JL acknowledges support by the Faculty of the European Space
Astronomy Centre.
DFT further acknowledges the Chinese Academy of Sciences visiting professorship program 2013T2J0007.
We thank M. Massi for facilitating us
her collected radio data and A. Papitto \& N. Rea for comments.

\end{document}